\documentclass[conference]{IEEEtran}
\IEEEoverridecommandlockouts
\usepackage{cite}
\usepackage{amsmath,amssymb,amsfonts}
\usepackage{algorithmic}
\usepackage{graphicx}
\usepackage{textcomp}
\usepackage{url}
\usepackage{xcolor}
\usepackage{caption}
\def\BibTeX{{\rm B\kern-.05em{\sc i\kern-.025em b}\kern-.08em
    T\kern-.1667em\lower.7ex\hbox{E}\kern-.125emX}}
\begin{document}
\title{IndieFake Dataset: A Benchmark Dataset for Audio Deepfake Detection}




\author{\IEEEauthorblockN{Abhay Kumar}
\IEEEauthorblockA{\textit{Dept. of Electrical Engineering} \\
\textit{Indian Institute of Technology Ropar}\\
Ropar, India \\
a3141b@gmail.com}
\and
\IEEEauthorblockN{Kunal Verma}
\IEEEauthorblockA{\textit{Dept. of Electrical Engineering} \\
\textit{Indian Institute of Technology Ropar}\\
Ropar, India \\
kunal.v.0975@gmail.com}
\and
\IEEEauthorblockN{Omkar More}
\IEEEauthorblockA{\textit{Dept. of Electrical Engineering} \\
\textit{Indian Institute of Technology Ropar}\\
Ropar, India \\
moreomkar4842@gmail.com}
}

\maketitle
\begin{abstract}
Advancements in audio deepfake technology offers benefits like AI assistants, better accessibility for speech impairments, and enhanced entertainment. However, it also poses significant risks to security, privacy, and trust in digital communications. Detecting and mitigating these threats requires comprehensive datasets. Existing datasets lack diverse ethnic accents, making them inadequate for many real-world scenarios. Consequently, models trained on these datasets struggle to detect audio deepfakes in diverse linguistic and cultural contexts such as in South-Asian countries. Ironically, there is a stark lack of South-Asian speaker samples in the existing datasets despite constituting a quarter of the worlds population. This work introduces the IndieFake Dataset (IFD), featuring 27.17 hours of bonafide and deepfake audio from 50 English-speaking Indian speakers. IFD offers balanced data distribution and includes speaker-level characterization, absent in datasets like ASVspoof21 (DF). We evaluated various baselines on IFD against existing ASVspoof21 (DF) and In-The-Wild (ITW) datasets. IFD outperforms ASVspoof21 (DF) and proves to be more challenging compared to benchmark ITW dataset. The complete dataset, along with documentation and sample reference clips, is publicly accessible for research use\footnote{Project Website: \url{https://indie-fake-dataset.netlify.app}}.
\end{abstract}

\begin{IEEEkeywords}
Audio Deepfake Detection, text-to-speech, Audio
\end{IEEEkeywords}

\section{Introduction and Related Work}

The rapid advancements in speech synthesis, particularly driven by deep learning, have enabled a wide range of applications, including personal assistants \cite{pa_siri}, audiobook narration \cite{tacotron}, and voice synthesis for individuals with disabilities \cite{pa_blind}\cite{pa_old_blind}. While these applications offer significant benefits, the rise of deepfakes introduces new risks, such as impersonation, misinformation, fraud, and privacy invasion. 

Audio deepfakes are primarily generated through Text-To-Speech (TTS) \cite{tacotron} and Voice Conversion (VC) \cite{wavenet_vc}. TTS technology converts text into spoken words, creating synthetic speech that closely mimics a target voice \cite{tacotron}. VC, on the other hand, transforms one person's voice to resemble another's while maintaining the linguistic content \cite{stargan_vc}. TTS models are typically categorized into three types: Concatenative \cite{concatenative}, Parametric \cite{parametric_tts}, and Deep-learning-based approaches \cite{tacotron}. Notable deep-learning approaches include: (i) Sequence-to-Sequence models such as Tacotron \cite{tacotron} and Tacotron2 \cite{tacotron2}, which map text sequences to spectrograms; (ii) Generative Adversarial Networks (GANs) like WaveGAN \cite{wavegan}, ParallelGAN \cite{parallelgan}, and MelGAN \cite{melgan}, which generate highly realistic audio samples; and (iii) Variational Autoencoders (VAEs) such as VITS \cite{vits}, EfficientTTS 2 \cite{efficienttts2}, and AutoVC \cite{autoVC}, which enhance diversity and quality in speech synthesis.

The proliferation of deepfakes has made it essential to develop robust detection methods to mitigate the risks posed by them. Various methods have emerged, such as neural networks for analyzing both pre-processed \cite{lcnn}, \cite{mesonet} and raw audio data \cite{rawnet3} to classify samples as either ``deepfake" or ``bonafide." Techniques for feature extraction, including MFCC \cite{mfcc_1995}, Mel-Spectrogram \cite{mel_spec}, Whisper Features \cite{whisper}, CQCC \cite{asvspoof21}, and LFCC \cite{mfcc_lfcc}, have also played a crucial role.

Recent advancements in Audio Deepfake Detection (ADD) \cite{asvspoof21}\cite{vicomtech}\cite{mis-avoidd} highlight promising progress. Moreover, complex-valued networks for voice anti-spoofing have emerged as a promising area within ADD research \cite{complex_valued_networks}. Further, several datasets, including LJSpeech \cite{ljspeech}, In-The-Wild \cite{in_the_wild}, ASVspoof21 (DF) \cite{asvspoof21}, FakeOrReal \cite{fake-real}, LibriTTS \cite{libritts}, VCTK \cite{vctk}, and VCC-2018 \cite{vcc2018}, have contributed significantly to the development of deepfake detection techniques. 

Among the existing datasets, ASVspoof21 (DF) is considered the largest and well-known dataset but lacks ethnic and linguistic diversity, particularly for Indian speakers. 
Moreover, its highly imbalanced nature (with 95 deepfake instances for every 5 bonafide ones) and speaker-independent setup limit model generalizability. The In-The-Wild (ITW) dataset, while being more balanced and speaker-dependent, also fails to include Indian speakers or address critical scenarios, such as content transfer between real and cloned voices, which are key to real-world security risks. This absence of datasets tailored to Indian speakers and scenarios limits the effectiveness of detection models in this context.

To address these limitations, we introduce the IndieFake Dataset (IFD), which includes 50 English-speaking Indian speakers and covers diverse genders, age groups, and backgrounds. The dataset contains 11.3 hours of genuine audio samples and 15.82 hours of deepfake audio samples, offering a valuable benchmark for detecting deepfakes in Indian-specific contexts. Our contributions include: 
\begin{itemize}
\item Introducing the IndieFake Dataset (IFD), which encompasses various challenges, as discussed in Section \ref{dataset_scenarios}, and features a subject-independent train-test split. 
\item Evaluating several baseline architectures for ADD on multiple datasets, including proposed IFD, ASVspoof21 (DF) \cite{asvspoof21}, and ITW \cite{in_the_wild}. 
\item Comparing the performance of existing datasets (ASVSpoof21 (DF) and ITW) against IFD using baseline methods across multiple scenarios. 

\end{itemize}

\section{IndieFake Dataset - Description}\label{data_description}


\subsection{Generation Methodology}\label{dataset_scenarios}
IFD covers four scenarios for data collection and generation in order to induce challenges in the dataset. Each scenario is discussed below:

\subsubsection{Deepfakes with hypothetical transcripts}  
In this scenario, deepfakes have been generated using hypothetical transcripts. For instance, we used an actor's voice to deliver a lecture on astrophysics, used a businessman's voice to recite literary works, etc. This scenario introduces a broad range of speech pattern, tones, and textual input into our dataset. A total of 2,562 deepfakes are generated under this scenario. 

\begin{figure}
    \centering
    \includegraphics[width=1\linewidth]{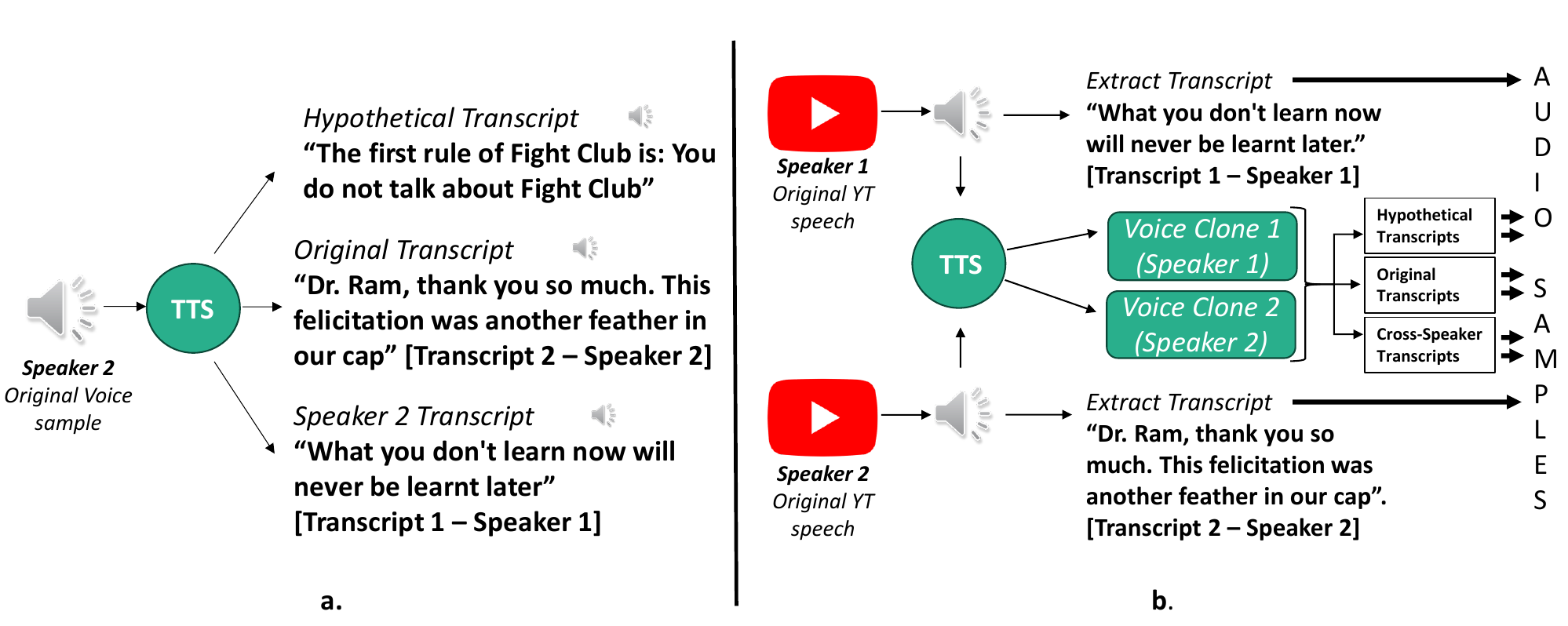}
    \vspace{-6mm}
    \caption{(a) Illustrates the scenarios that are considered for deepfake generation. (b) Illustrates the pipeline for deepfake audio generation and bonafide audio collection.}
    \label{data_plan}
\end{figure}

\subsubsection{Deepfakes with transcript of the same speaker}
Deepfakes for a speaker are generated using the speaker's transcript. In this scenario, the synthesized audio closely aligns with the speaker's speech patterns and content. This poses a significant challenge for audio verification systems, as deepfakes of verified users can be a serious threat. A total of 1,245 such deepfakes were generated in this scenario.

\subsubsection{Deepfakes with transcript of another listed speaker} 
Deepfakes for a speaker are generated using transcripts of another speaker listed in the dataset. This enables us to study content transfer across different speakers, which can have implications for speech synthesis, and voice conversion. This scenario introduces a challenge in the dataset as the model needs to distinguish between genuine samples and cross-speaker deepfakes with identical content. A total of 4,667 such deepfakes are generated under this scenario.

8,474 deepfake audio samples are created across the three scenarios, and this number increased to 11,396 using audio augmentation techniques. 

\begin{figure}[h]
    \centering
    \includegraphics[width=1\linewidth]{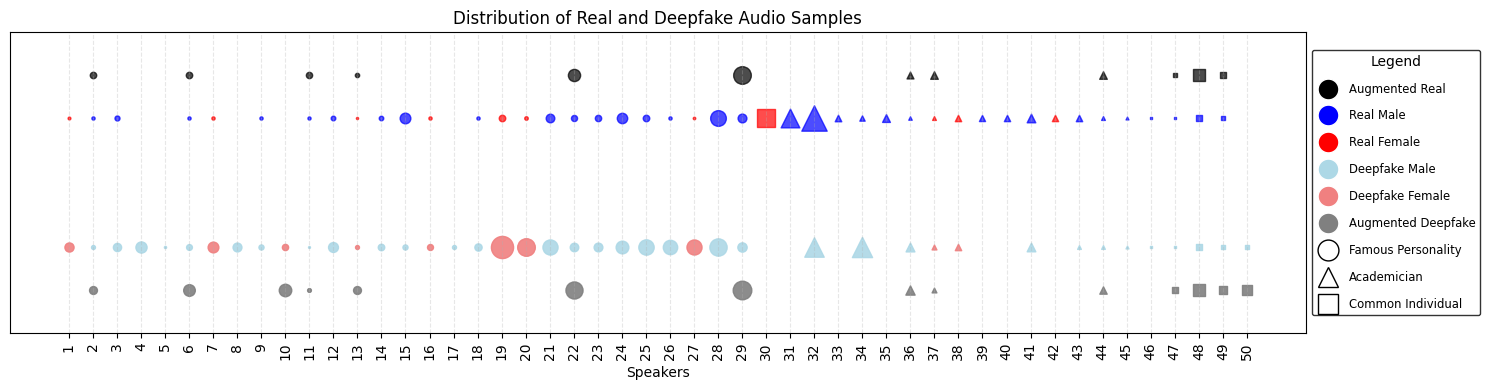}
    \caption{Distribution of audio samples for each speaker in the dataset. Each vertical line represents a speaker and the size of the blob is proportional to the number of audio samples. From top to bottom, the first row represents augmented bonafide samples, the second row represents bonafide samples, the third row represents deepfake samples, and the fourth row represents augmented deepfake samples. The blobs are color-coded as follows: blue for bonafide male, red for bonafide female, cyan for deepfake male, pink for deepfake female, black for augmented bonafide samples, and grey for augmented deepfake samples. Circle represents famous personalities, triangle represents academicians, and square represents common individuals.
    }
    \label{fig:datasetrepresentation}
\end{figure}



\subsubsection{Bonafide Samples}
Bonafide audio samples are sourced from YouTube videos licensed under Creative Commons. IFD includes a variety of public figures as well as common individuals. We used public interviews and speeches to collect audio samples of public figures and voiced sample transcripts to collect audio samples of common individuals.
We also applied various augmentation techniques, including Gaussian noise \cite{gaussian_noise}, background noise \cite{back_noise}, short noises \cite{short_noises}, reverse \cite{rev}, pitch shifting \cite{pitch_shift}, time masking \cite{time_masking}, and impulse response \cite{impulse_response}. All audio samples of a selected speaker are augmented using four randomly selected techniques mentioned above. 6,374 bonafide audio samples are collected under this scenario which are increased to 7,164 audio samples by using audio augmentation techniques. Additionally, we incorporated a set of miscellaneous audio samples that are not associated with any specific speaker to introduce variability in the dataset. For this purpose we used \cite{nptel_github}, which brought the final count of bonafide samples to 8,164.

\begin{table}[t]
    \centering
    \begin{tabular}{|l|c|c|c|c|}
        \hline
        \textbf{Type of Data} & \textbf{Speaker} & \textbf{Bonafide} & \textbf{Deepfake} & \textbf{Total} \\
        \hline
         & Female & & & \\
        \hline
        Complete & Speaker-13
        & 45 & 150 & 195 \\
        Bonafide Only & Speaker-45
        & 606 & 0 & 606 \\
        Deepfake Only & Speaker-20
        & 0 & 596 & 596 \\
        \hline
        \multicolumn{2}{|c|}{\textbf{Total (Female)}} & 651 & 746 & 1397 \\
        \hline
        \hline
         & Male & & & \\
        \hline
        Complete & Speaker-23
        & 209 & 311 & 520 \\
        Deepfake Only & Speaker-50
        & 0 & 195 & 195 \\
        Bonafide Only & Speaker-22
        & 360 & 0 & 360 \\
        Deepfake Only & Speaker-6
        & 0 & 325 & 325 \\
        Complete & Speaker-21
        & 144 & 439 & 583 \\
        Bonafide Only & Speaker-28
        & 475 & 0 & 475 \\
        Deepfake Only & Speaker-4
        & 0 & 239 & 239 \\
        \hline
        \multicolumn{2}{|c|}{\textbf{Total (Male)}} & 1188 & 1509 & 2697 \\
    \hline
    \end{tabular}
    \vspace{1mm}
    \caption{Speakers in Testing Set}
    \label{tab:dataset_split1}
\end{table}

Dataset construction involves these steps: First, an Indian speaker is selected, and their English audio is extracted from YouTube videos. For deepfake samples, a portion of the extracted audio is used to train a TTS model with softwares like Amazon Polly, Play.ht, or ElevenLabs, as shown in Fig. \ref{data_plan} (b). Deepfake samples are generated using these TTS models and speaker transcripts, under one or more of the three deepfake scenarios discussed above, as illustrated in Fig \ref{data_plan} (a). The pipeline for dataset creation is demonstrated in Fig \ref{data_plan} (b). All audio samples have an average length of 5 seconds.

\begin{table*}[t]
\centering
\begin{tabular}{|l|rrrr|rrrr|rr|rr|rr|}
\hline
    Setup $\rightarrow$         
    & \multicolumn{4}{c|}{T(IFD)(80\%) + V(IFD) (20\%)}                        
    & \multicolumn{4}{c|}{T(IFD)(80\%) + V(ITW)}                    
    & \multicolumn{2}{c|}{T(ASV)}         
    & \multicolumn{2}{c|}{T(ASV)}                      
    & \multicolumn{2}{c|}{T(IFD)(80\%)} \\ 
\hline
    Condition $\rightarrow$       
    & \multicolumn{2}{c|}{Scratch (A)}                              
    & \multicolumn{2}{c|}{P(ASV) (B)}     
    & \multicolumn{2}{c|}{Scratch (C)}                              
    & \multicolumn{2}{c|}{P(ASV) (D)}           
    & \multicolumn{2}{c|}{V(ITW) (E)}                
    & \multicolumn{2}{c|}{V(IFD) (F)}                       
    & \multicolumn{2}{c|}{V(ASV) (G)} \\ 
\hline
    Model $\downarrow$
    & \multicolumn{1}{l|}{Test}   & \multicolumn{1}{l|}{EER}    
    & \multicolumn{1}{l|}{Test}   & \multicolumn{1}{l|}{EER} 
    & \multicolumn{1}{l|}{Test}   & \multicolumn{1}{l|}{EER}    
    & \multicolumn{1}{l|}{Test}   & \multicolumn{1}{l|}{EER} 
    & \multicolumn{1}{l|}{Test}   & \multicolumn{1}{l|}{EER} 
    & \multicolumn{1}{l|}{Test}   & \multicolumn{1}{l|}{EER} 
    & \multicolumn{1}{l|}{Test}   & \multicolumn{1}{l|}{EER} \\ 
\hline
    LFCC-LCNN      
    & \multicolumn{1}{r|}{85.64}  & \multicolumn{1}{r|}{0.119} 
    & \multicolumn{1}{r|}{64.061} & {0.2658}                   
    & \multicolumn{1}{r|}{69.73}  & \multicolumn{1}{r|}{0.372} 
    & \multicolumn{1}{r|}{\textbf{65.117}} & {0.3362}                  
    & \multicolumn{1}{r|}{\textbf{76.36}} & \textbf{0.233}                    
    & \multicolumn{1}{r|}{46.311} & \textbf{0.244}                    
    & \multicolumn{1}{r|}{54.708}   & {0.395} \\ 
\hline
    MFCC-LCNN      
    & \multicolumn{1}{r|}{89.47}  & \multicolumn{1}{r|}{0.100} 
    & \multicolumn{1}{r|}{59.552} & {0.271}                   
    & \multicolumn{1}{r|}{65.365} & \multicolumn{1}{r|}{0.389}  
    & \multicolumn{1}{r|}{63.824} & {0.424}                  
    & \multicolumn{1}{r|}{67.99} & {0.34}                     
    & \multicolumn{1}{r|}{43.137} & {0.39}                     
    & \multicolumn{1}{r|}{53.041} & {0.442}  \\ 
\hline
    LFCC-Mesonet   
    & \multicolumn{1}{r|}{87.64}  & \multicolumn{1}{r|}{0.116} 
    & \multicolumn{1}{r|}{44.785} & {0.250}                   
    & \multicolumn{1}{r|}{74.03}  & \multicolumn{1}{r|}{0.282} 
    & \multicolumn{1}{r|}{62.833} & {0.482}                  
    & \multicolumn{1}{r|}{65.3}  & {0.439}                    
    & \multicolumn{1}{r|}{38.638} & {0.505}                    
    & \multicolumn{1}{r|}{72.738} & \textbf{0.354}   \\ 
\hline
    MFCC-Mesonet   
    & \multicolumn{1}{r|}{78} & \multicolumn{1}{r|}{0.197} 
    & \multicolumn{1}{r|}{\textbf{71.339}} & {0.299}                 
    & \multicolumn{1}{r|}{66.63}  & \multicolumn{1}{r|}{0.334}  
    & \multicolumn{1}{r|}{61.681} & {0.420}                 
    & \multicolumn{1}{r|}{60.42} & {0.698}                    
    & \multicolumn{1}{r|}{38.674} & {0.469}                   
    & \multicolumn{1}{r|}{\textbf{82.936}} & {0.358}   \\ 
\hline
    Rawnet3        
    & \multicolumn{1}{r|}{\textbf{94}} & \multicolumn{1}{r|}{\textbf{0.061}} 
    & \multicolumn{1}{r|}{68.835} & \textbf{0.130}                  
    & \multicolumn{1}{r|}{\textbf{77.01}}  & \multicolumn{1}{r|}{\textbf{0.21}}   
    & \multicolumn{1}{r|}{59.26}    & \textbf{0.335}                    
    & \multicolumn{1}{r|}{57.99} & {0.497}                    
    & \multicolumn{1}{r|}{\textbf{57.001}} & {0.402}                  
    & \multicolumn{1}{r|}{68.033}   & {0.462}  \\ 
\hline
\end{tabular}
\vspace{1mm}
\caption{Experimentation Results. T represents Training, V represent Validation and P represents using Pre-trained weights. 
Column A represents results of baselines when trained on IFD from scratch and tested on IFD, 
Column B represents results of baselines when finetuned on IFD using pre-trained weights from ASVspoof21 (DF) and tested on IFD, 
Column C represents results of baselines when trained on IFD from scratch and tested on ITW,
Column D represents results of baselines when finetuned on IFD using pre-trained weights from ASVspoof21 (DF) and tested on ITW,
Column E represents results of baselines when trained on ASVspoof21 (DF) and tested on ITW,
Column F represents results of baselines when trained on ASVspoof21 (DF) and tested on IFD, and
Column G represents results of baselines when trained on IFD and tested on ASVspoof21 (DF).
}
\label{experiments_results}
\end{table*}

\subsection{Dataset Details}
IndieFake Dataset (IFD) is a subject dependent dataset containing bonafide and deepfake audio samples of various speakers, speaking English, from different regions of India. IFD features 50 speakers from different backgrounds (Celebrities, Politicians, Academicians, Businessmen, and Common Public), age groups (18-75 years), regions and accent. All bonafide audios are taken from YouTube, licensed under Creative Commons. We have maintained a balance in the dataset over speaker background and the number of deepfake and bonafide samples. This dataset contains 8,164 (around 11.3 hours) bonafide audio samples and 11,396 (around 15.82 hours) deepfake audio samples of an average duration of 5 seconds. Refer to Fig.\ref{fig:datasetrepresentation} for more details.

We observe from Fig. \ref{fig:datasetrepresentation}, speaker number 4, 5, 8, 10, 17, and 50 only contain deepfake audio samples in the dataset, speaker number 30, 31, 33, 35, 39, 40, and 42 only contain bonafide audio samples in the dataset, rest of the speakers have both deepfake and bonafide audio samples.

\subsubsection{Train-Test Split}
We adopted subject independent splitting approach to mitigate subject dependency and test the generalization capability of models. Subject independency is maintained by ensuring no speaker's bonafide or deepfake samples appear in both training and testing sets. The dataset is split into 80\%:20\% train:test split. IFD contains 38 male speakers and 12 female speakers. The male-to-female and bonafide-to-deepfake ratios in the train and test splits are consistent with the overall dataset. The testing set contains total of 4094 audio samples out of which 1839 are bonafide and 2255 are deepfakes while the training set contains a total of 15466 audio samples out of which 6327 are bonafide and 9139 are deepfakes.

While making the testing set we have considered four scenarios: (i) Only Deepfake, these speakers only have deepfakes in the dataset and all the samples are taken in the testing set, e.g. in Fig. \ref{fig:datasetrepresentation} Speaker 50, (ii) Only Bonafide, these speakers only have bonafide samples in the dataset and all the samples are taken in the testing set, e.g. in Fig. \ref{fig:datasetrepresentation} Speaker 40 (iii) Bonafide and Deepfake, had both bonafide and deepfakes in the dataset and all the samples are taken in the testing set, e.g. in Fig. \ref{fig:datasetrepresentation} Speaker 13 and (iv) Bonafide or Deepfake, similar to (iii) but only either of the class has been taken in the testing set, e.g. in Fig. \ref{fig:datasetrepresentation} Speaker 6.  


\section{Evaluation}
This section discusses the diversity and representativeness of proposed IFD dataset in comparison to ASVspoof21 (DF) and ITW dataset using various baselines. Further, we discuss the results and ablation study.

\subsection{Baseline Methods}
Audio deepfake detection methods are generally categorized into i) front-end-based models, and ii) raw audio signal based models. Front-end based methods use pre-processing techniques for audio transformation such as LFCC, MFCC and Mel-Spectograms. Whereas, raw audio signal-based models are end-to-end deep learning approaches. The pre-processing techniques used for front-end-based baseline models are MFCC and LFCC. We used LCNN \cite{lcnn} and Mesonet \cite{mesonet} as front-end-based baseline models while Rawnet3 \cite{rawnet3} is used as end-to-end deep learning baseline model. The details of baselines are illustrated in Table \ref{baseline_models}.

    
   


\begin{table}[h]
    \centering
    \begin{tabular}{|l|l|l|}
        \hline
        \textbf{Model} & \textbf{Description} & \textbf{Acoustic} \\
        & & \textbf{Features}\\
        \hline
        LCNN \cite{lcnn} & Combination of convolutional & MFCC, LFCC \\
        & layers and Max Feature Map activation & \\
        MesoNet \cite{mesonet} & Originally used in detecting & MFCC, LFCC \\
        & facial video deepfakes & \\
        RawNet3 \cite{rawnet3} & Uses parameterized filter bank along & Raw audio \\
        & with bottleneck residual blocks & signal\\
        \hline
    \end{tabular}
    \vspace{0.5mm}
    \caption{Baseline Models considered for experimental analysis}
    \label{baseline_models}
\end{table}


\subsection{Datasets}
\begin{itemize}
    \item \textbf{ASVspoof21 (DF):} The ASVspoof21 (DF) dataset \cite{asvspoof21} is a benchmark dataset designed for research in automatic speaker verification (ASV) and spoofing detection. It aims to advance the SOTA in detecting spoofing attacks in speaker verification systems. ASVspoof21 (DF) contains around 20,637 bonafide audio samples and 572,616 deepfake audio samples.

    \item \textbf{In-The-Wild (ITW):} This dataset \cite{in_the_wild} is generally used for evaluation purposes. ITW contains around 38 hours of audio clips which have been labelled as deepfake (around 17 hours) and bonafide (around 21 hours). The dataset is in English language, with an average audio sample length of 4.3 seconds.
\end{itemize}

\subsection{Experimental Setup}
Each sample has been stored in WAV format and is loaded at frequency of 16KHz. To standardize audio samples to 5 seconds, we trimmed any exceeding length and padded shorter ones by repetition. In our experiments, we used 125,000 randomly selected audio samples from ASVspoof21 (DF) for training and pre-training in different scenarios. We also used pre-trained weights \cite{kawa} trained in similar way on ASVspoof21 (DF) for various baselines. All baseline models are trained for 10 epochs on the ASVspoof21 (DF) dataset and 50 epochs on the IFD dataset. For front-end-based baselines learning rate, $\eta = 10^{-4}$ and weight decay $\theta = 10^{-4}$, while for Rawnet3 $\eta = 10^{-3}$ and $\theta = 5 \times 10^{-5}$. We used Adam optimizer, binary cross-entropy loss and an average batch size of 64 for all baselines. The metrics used are Testing Accuracy and Equal Error Rate (EER) \cite{asvspoof21}. Lower EER is preferred.

\subsection{Results}
This section provides a comprehensive study assessing the performance of baselines after training on IFD and ASVspoof21 (DF) across different cross-validation scenarios with IFD, ASVspoof21 (DF) and ITW, in Table \ref{experiments_results} and Table \ref{Dataset_results}. It can be observed from Table \ref{Dataset_results} that the EER reported after training on IFD is consistently better compared to EER reported after training on ASVspoof21 (DF) for most baselines. This indicates that the representativeness and diversity of the proposed IFD are better compared to ASVspoof21 (DF), despite the smaller size of the proposed IFD (one-sixth of the size of ASVspoof21 (DF)). This leads to a significant reduction in training time and computational resource utilization, while delivering better results, showcasing the dataset's high quality and effectiveness.

\begin{table}[h]
    \centering
    \begin{tabular}{|l|c|c|}
        \hline
        \textbf{Dataset} $\rightarrow$ & \textbf{ASVspoof21 (DF)} & \textbf{IFD} \\
        \hline
        Baseline $\downarrow$ & EER & EER \\
        \hline
        LFCC-LCNN & \textbf{0.233} & 0.3718 \\
        MFCC-LCNN & \textbf{0.34} & 0.389 \\
        LFCC-MesoNet & 0.439 & \textbf{0.2819} \\
        MFCC-MesoNet & 0.698 & \textbf{0.334} \\
        RawNet3 & 0.497 & \textbf{0.21} \\
        \hline
    \end{tabular}
    \vspace{2mm}
    \caption{Comparison of EER when (i) trained on ASVspoof21 (DF) and tested on ITW and, (ii) trained on proposed IFD and tested on ITW}
    \label{Dataset_results}
\end{table}

Results in Table \ref{experiments_results}, columns A and B show 
that the performance deteriorates after using pretrained weights from ASVspoof21 (DF), a similar trend is observed in columns C and D. This decline in performance can be attributed to the highly imbalanced nature of ASVspoof21 (DF) dataset, where the ratio of deepfake to bonafide samples is roughly 95:5. Pretraining on ASVspoof21 (DF) generally hinders model performance, likely due to its highly imbalanced nature. To further support this argument, we conducted a cross-validation experiment shown in Table \ref{experiments_results} columns F and G. The results indicate that test accuracy in Column G is consistently better than in Column F, and the EER in Column G is superior in 2 out of 5 cases, with the remaining 3 cases showing comparable results. These findings also suggest the effectiveness of IFD dataset over ASVspoof21 (DF). Further, Table \ref{experiments_results} Columns E and F demonstrates that, in most cases, the EER is higher, and in all cases, the testing accuracy is consistently lower when tested on IFD compared to ITW after training on ASVspoof21 (DF). Poor results on IFD demonstrates its complexity, making it a better choice for a benchmark dataset. 

\subsection{Ablation Study}
\begin{figure}
    \centering
    \includegraphics[width=1\linewidth]{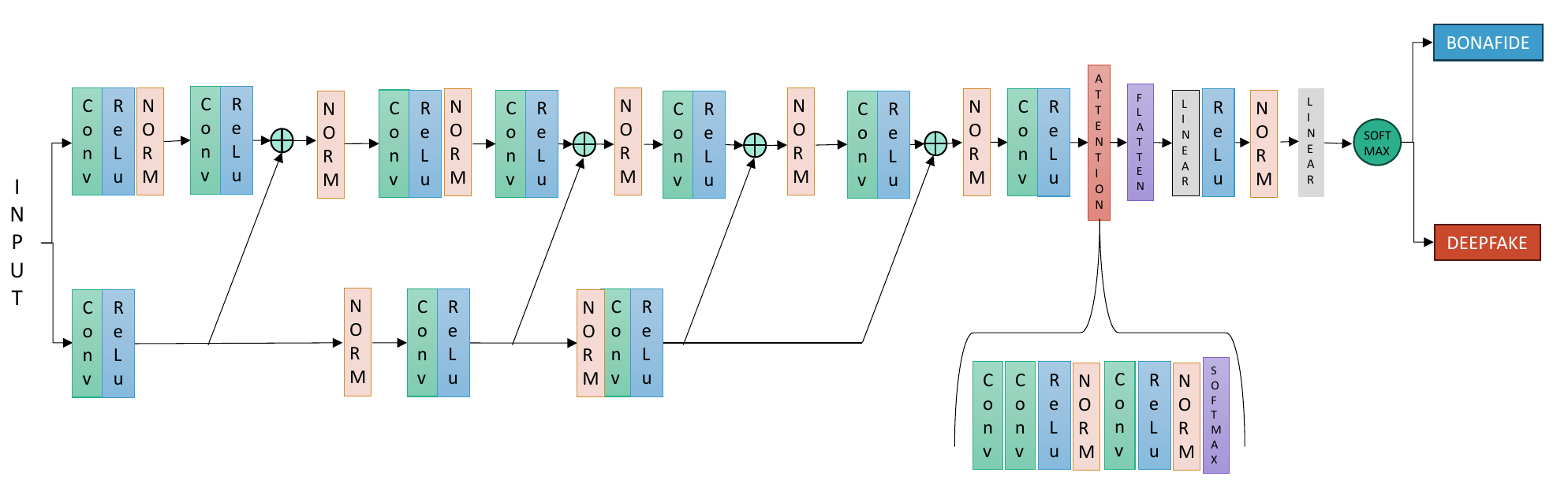}
    \caption{ICDD (Inter-Channel Deepfake Detection)}
    \label{fig:icdd}
\end{figure}

\begin{figure}
    \centering
    \includegraphics[width=1\linewidth]{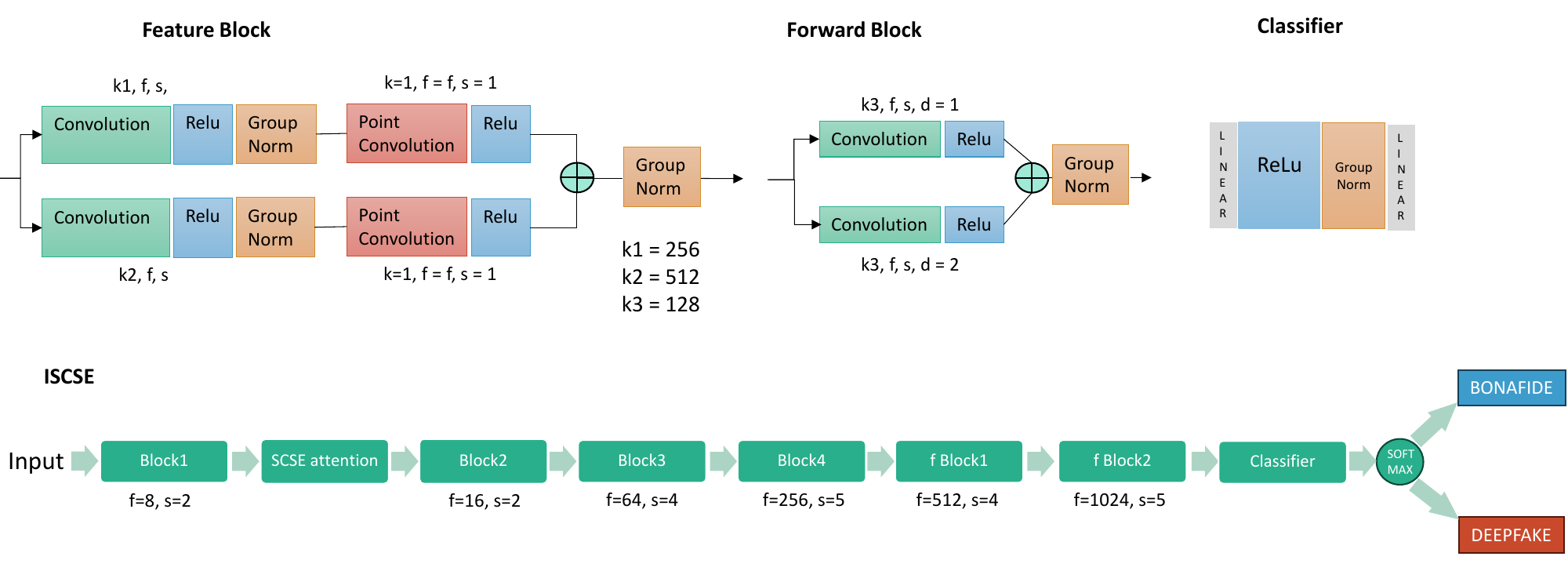}
    \caption{ISCSE (Inter Spatial-Channel Squeeze and Excitation)}
    \label{fig:iscse}
\end{figure}

In this section we discuss and evaluate various aspects of end-to-end deep-learning approach for deepfake audio detection. RawNet3 \cite{rawnet3} is a end-to-end deep-learning based model which was designed to compute speaker embedding but shows great results in ADD as well. RawNet3 uses parameterized filter-bank \cite{filter_banks}, which is trainable layer that converts raw audio into time-frequency representation, at its initial stage. Table \ref{ablation}, row 1 and 2 lists the performance of RawNet3 and a modification of RawNet3, where we replaced the F-bank layer by convolution blocks, respectively. RawNet3 greatly outperforms RawNet3 modification indicating the importance of the parameterized filter-bank layer.
We also investigated the performance of complete end-to-end deep-learning models on raw audio samples. For these experiments we used two convolutional models, (i) ICDD (Inter-Channel Deepfake Detection) \ref{fig:icdd} and (ii) ISCSE (Inter Spatial-Channel Squeeze and Excitation) \ref{fig:iscse}. ICDD uses multi-channel convolutional blocks for feature extraction and also uses SCSE attention block \cite{scse}. ISCSE also uses multi-channel convolutional blocks for feature extraction with some modifications and an SCSE attention block.

Through our experiments, we observed that raw audio inputs pose challenges due to large variations caused by factors such as loudness, noise, and pitch. This makes normalization crucial in deep learning models that utilize raw audio, directly impacting the robustness of the solution. However, batch normalization, available in popular deep learning frameworks, is not used during testing. As a result, models that perform well during training may exhibit poor performance during testing. To address this issue, deep learning models like RawNet3 \cite{rawnet3} incorporate custom normalization blocks to ensure consistent performance.

We evaluated the performance of Group Normalization \cite{groupnorm} and Instance Normalization \cite{instanceNorm}, which are alternative normalization techniques. Table \ref{ablation}, row 3, 4, and 5, lists the performance of ISCSE using Group Normalization with group sizes of 8, 4, and 2, respectively. The results indicate that model performance improves with smaller group sizes, peaking at a group size of 2. Additionally, Table \ref{ablation}, row 6 and 8, presents the performance of ICDD with Group Normalization and Instance Normalization, respectively. Furthermore, Table \ref{ablation}, row 7 and 9, displays the performance of an ICDD modification that includes an LSTM layer with Group Normalization and Instance Normalization, respectively. We can observe that the performance of Group Normalization is better compared to Instance Normalization.   

\begin{table}[h]
\centering
\begin{tabular}{|c|c|c|c|c|}
\hline
Sr. No. & Model   & Variation           & Test Accuracy & EER   \\ \hline
1 & RawNet3 & -                   & 94            & 0.061 \\ \hline
2 & RawNet3 & W/O F-Bank          & 55.126        & 0.131 \\ \hline
3 & ISCSE   & GN = 8       & 63.352        & 0.353 \\ \hline
4 & ISCSE   & GN = 4       & 69.826        & 0.308 \\ \hline
5 & ISCSE   & GN = 2       & 76.667        & 0.276 \\ \hline
6 & ICDD    & GN = 4         & 74.566        & 0.275 \\ \hline
7 & ICDD    & LSTM + GN = 4  & 72.938        & 0.304 \\ \hline
8 & ICDD    & IN       & 55.241        & 0.464 \\ \hline
9 & ICDD    & LSTM + IN & 59.156        & 0.435 \\ \hline
\end{tabular}
\caption{Ablation Studies: GN here means Group Normalization and IN means Instance Normalization}
\label{ablation}
\end{table}

\section{Conclusion}\label{conclusion}
In this paper, we present a new dataset specifically targeting ADD for Indian applications. Despite its smaller size, the IndieFake Dataset (IFD) demonstrates superior performance as a training resource compared to the renowned ASVSpoof21 (DF) dataset. Our experiments further show that IFD presents a more challenging benchmark compared to the In-The-Wild (ITW) dataset, establishing it as a more effective resource for evaluating ADD systems in Indian contexts.


\end{document}